# SHOW YOUR WORK: AN EXPLORATORY STUDY OF STUDENT EXPERIENCES OF TURNITIN CLARITY

A PREPRINT

Jasper Roe [1*], Mike Perkins [2]

[1] School of Education, Durham University, United Kingdom

[2] Centre for Research and Innovation, British University Vietnam, Vietnam

[*] Corresponding Author: jasper.j.roe@durham.ac.uk

September 2026

## Abstract

Process-capture platforms promise to make the writing behind an assessed artefact visible, and vendors increasingly market them as a fairness measure rather than a detection tool. Their acceptability to the students who must write within them, however, is largely untested. This exploratory qualitative study examines how students' expectations of one such platform, Turnitin Clarity, compared with their experience of using it. Seven students at a UK university, given no onboarding to the platform, took part in pre-use focus groups. Six then completed an unassessed 500-word writing task over five days, before all seven returned for a post-use focus group. Data was analysed using reflexive thematic analysis, with Expectation-Confirmation Theory as an orienting frame.

The overall pattern was one of mixed confirmation. Participants' functional expectations were largely met, and the sense of being watched that they had anticipated persisted after use: for some it was offset by perceived fairness, while for others it heightened self-monitoring. The bounded AI assistant divided them: its limits were welcomed for marking out acceptable use, but some felt it weakened their ownership of the work and worried it would flatten what they produced. Across both phases, participants weighed the costs of observation against perceived gains in fairness and asked for transparency to run in both directions. Several described already writing defensively in anticipation of accusations of AI misuse. Acceptance of process technologies was reported as being conditional on practice time, two-way transparency, and explicit data governance. Participants also questioned whether a single linear document can represent a writing process they described as messy and multimodal. Across these accounts, process capture did not sit outside the writing it recorded but reorganised the practice it set out to observe.

**Keywords:** generative artificial intelligence; process data; academic integrity; assessment design; student perceptions

# Introduction

When learners create digital artefacts for educational assessments, the process of creation is often hidden. Product-oriented assessment of this kind does not provide transparency into the process of creating the final product (Fawns et al., 2026). The disruptive effects of new technologies, including Generative AI (GenAI), have in turn made it more challenging to ascertain whether such products are representative of a student's work, knowledge, or ability. Multiple approaches have been posited to respond to this problem, ranging from a reduction in the use of technology (such as returning to pen and paper examinations), to increasing the application of technology in assessment (for example, using detection software). At the same time, several products have been developed by educational technology and integrity product providers, which focus on providing assurance of learning and producing a faithful representation of students' learning process and production of an assessment artefact.

Process-capture platforms offer several potential solutions to this issue, such as ensuring that assessments are fairer (i.e., students who have access to better AI technologies will not be able to receive better grades), and providing the assurance of academic integrity needed to make inferential claims about assessment outcomes (i.e., student results are accurate representations of their abilities and knowledge). On the other hand, these platforms carry with them risks of increasing surveillance, while their ability to assure academic integrity is as yet unproven. Other risks emerge in maintenance of validity when using process assessment more generally; Fawns et al. (2026) argue that process assessment risks becoming "a security mechanism for product-based tasks (e.g. monitoring technology use or authenticating identity), rather than treating process as evidence of learning" (p. 7). This in turn can lead to an increase in performative writing which damages the validity of the assessment.

Turnitin Clarity is a new process-capture platform that also includes a bounded AI assistant for student use. There is a need for understanding the implications, efficacy, and suitability of platforms like these in educational assessment. As a first step, we begin by exploring student perspectives on this platform. The rationale for this is that students are the stakeholders whose writing practice these platforms will capture and use. As a result, their willingness to use them, their perceptions and experience are a key condition of viability. Through pre- and post-use focus groups oriented around adapted Expectation-Confirmation Theory (ECT) dimensions, we examined how seven students' expectations of Turnitin Clarity compared with their accounts after a writing task in the platform. We used the following research questions:

1. How do university students' pre-use expectations of an AI-enabled, process-tracking writing platform compare with their views following an experience of the platform?
2. What do students' views reveal about the acceptability of these platforms in higher education assessment?

The analysis shows mixed confirmation: the anticipated sense of surveillance persisted after use, offset for some by perceived gains in fairness, while acceptance remained conditional on practice time, two-way transparency, and explicit data governance. Participants also questioned whether a linear record could accommodate non-linear writing practices. Across these accounts, process capture did not observe writing from the outside. Participants were already writing defensively before they engaged with the platform, and for some, writing within it further altered what they wrote and how, so the record it produces is, in part, an artefact of the recording. The remainder of the paper situates process capture within post-GenAI assessment, explains the study design, presents the six themes developed, and discusses implications for assessment practice.

# Literature Review

## Process Data as Research

Process data does not have a widely agreed upon definition, but Anghel et al. (2024) define it in relation to computer-delivered educational assessments as data that captures response times, keystrokes, selections and navigation, among others. It is important to note that this definition does not account for process data that may occur in other forms of education, for example the completion of an art portfolio using different mediums, or the production of components in a physical product prototype. This falls outside of the scope of the current work, and so in what follows, we focus on the idea of process data that is captured through various inputs and actions made on a device during the creation of digital textual assessment artefacts.

This form of digital process data has been used outside of the field of assessment for some time. Research in composition studies has historically made use of keylogging technology to generate insight into writing. Leijten and Van Waes (2013) note that the rationale underlying logging which keys and buttons are pressed, and when, enables an assessment of the cognition behind the output, while other researchers have developed sets of procedures to better examine the relationship between keystroke logs and cognition, focusing on the features such as pauses in typing, bursts of writing, and revisions (Baaijen et al., 2012). This is the epistemic foundation of exploring process data when created through computing.

## Process Data as an Answer for GenAI in Assessment

Multiple approaches have been proposed to deal with the "wicked problem" of GenAI and assessment (Corbin et al., 2026). One of the first methods which emerged as a response to the launch of ChatGPT was AI generated content (AIGC) detection technologies. However, many studies have demonstrated that these technologies are both technically imperfect and ethically problematic (Liang et al., 2023; Perkins, Roe, et al., 2024; Sun et al., 2026). Following this, methods turned to options such as the redesign of assessment focusing on the role of AI use (Perkins, Furze, et al., 2024; Perkins et al., 2025), the development of interrelated components that comprise more and less secure elements (Roe, Perkins, & Giray, 2026), and the creation of 'lanes' of secured and non-secured assessment types (Bridgeman et al., 2024). Studies on teachers have also shown a preference for a return to written or oral examinations (Bower et al., 2024).

Multiple authors have also suggested ways of including a greater focus on process rather than product data. Black Box Assessment has been proposed as a model which enables capturing and assessing process and cognitive development, as well as recognising the value of errors as formative indicators (Winstone et al., 2026). In the field of TESOL studies, Process Pedagogy 2.0 has been proposed as a model which values intellectual and behavioural activities (for example refining and evaluating output) rather than merely assessing linguistic markers (Elturki, 2026). There are no simple solutions to the problems posed by AI and assessment, and this burden has been labelled as one that resists singular 'answers', but must be dealt with through iterative negotiation, compromise, and adaptation (Corbin et al., 2026). A further challenge is scale: human-centred approaches to verifying authorship, such as structured dialogue with students, may become impractical in large cohorts unless process evidence is built into assessment design by default (Ucan, 2026).

## Process Data as Surveillance

The capture of student data in written assessment has a history of being viewed through the lens of surveillance and discipline. Zwagerman (2008) highlights that plagiarism detection can be viewed through the Foucauldian lens of the panopticon, while Jensen (2010) argues that process portfolios extensively discipline student-writers and may offer more chances for educators to enact disciplinary behaviours on them. Zwagerman (2008) contends that if academic honesty is used to justify surveillance, students may resist or act defiantly, with academic dishonesty being the primary means of resistance. Roe (2022) similarly reframes academic dishonesty as a struggle for intersubjective recognition, locating it in students' relationships with their institutions rather than in individual rule-breaking alone. Calderwood's (2026) research found that using digital surveillance mechanisms, if relied upon, can damage the relational trust between student and teacher, suggesting that such tools, if used, must be aligned with teachers' agency and educational values.

### Process Capturing Technologies in the Age of GenAI

There is little empirical research on process tracking software for assessment in the age of GenAI to date, with the work of Fernandes and McIntyre (2025) being the first that we were able to identify. Fernandes and McIntyre (2025) conducted a critical analysis of two platforms: Draftback and Grammarly Authorship. The authors note that these interfaces are adopted both by instructors wishing to ensure AI use does not happen in their courses, and independently by students who wish to proactively gather defensive data lest they be accused of AI misuse. The authors contend that these two platforms leverage fear to gain users, and that such surveillance is at odds with developing supportive classrooms, and that there may be cases when students should be given the ability to experiment with GenAI freely. A further empirical study deployed Grammarly Authorship in a large undergraduate course, analysing 310 essays to quantify word-level human and AI contributions; it demonstrated a classroom application of source tracking while also documenting attribution failures in the beta tool that was used (Harfe & Thomas, 2026).

Fernandes and McIntyre's (2025) critical interface analysis and Harfe and Thomas's (2026) quantitative deployment provide important early evidence, but neither examines students' lived experience of an integrated writing platform. There remains a need to study platforms that go beyond the capabilities of Draftback and Grammarly Authorship, especially as Draftback predates recent advances in GenAI.

Fernandes and McIntyre (2025) highlight three general features that process surveillance tools use. These include tracking the time students have spent within the document, identifying text that has been copied and pasted or written by the student, and the production of a recording or replay of the writing process. The platform under study in this work, Turnitin Clarity, offers these functionalities, but also goes further with the integration rather than exclusion of a GenAI tool. The platform contains a writing interface for students that has a bounded AI assistant, which can be aligned to a grading rubric. The chats with AI that are undertaken by the student are also optionally visible to the instructor, generating further evidence of the student's process (Turnitin, 2026). Turnitin states that the use of this platform helps create a level playing field for students by promoting transparency, as well as benefiting instructors by enabling them to "*view students' entire drafting process, including pasted text, writing time, construction time, and draft history. They can also access integrity insights, set AI usage policies for assignments, and provide tailored feedback*" (Turnitin, 2026).

## Methodology

In considering the above literature, composition research has established that keystroke and process data may give insight into writing cognition, while increasingly, process data capture is seen as a potential, or at least partial, solution to the problems posed by GenAI in assessment. Literature on process data suggests surveillance is a concern, but few studies have explored students' experience of products designed for the post-GenAI assessment landscape.

The study used a qualitative, longitudinal design, consisting of a pre-task focus group, a structured task using the Turnitin Clarity platform, and a post-task focus group. The design was exploratory, given that this is a novel platform. Ethics approval was granted by the institution's review board prior to the study's commencement. In the first phase, focus groups were used to explore participants' perceptions of the software, concerns, and beliefs about the nature of Turnitin Clarity. Following this, participants were given five days of access to the platform and asked to complete a writing task, making use of the functions as they wished.

After the writing task, the second focus group enabled a longitudinal comparison between participants' expected accounts and their post-use accounts. Focus groups were selected because their interactional format allows participants to develop, qualify, and contest accounts collectively, an established approach in educational technology and academic integrity research (Balida et al., 2024; Esposito, 2025; Kell et al., 2025; Perkins et al., 2026), including studies using Turnitin's technologies specifically (Buckley & Cowap, 2013). Recruitment occurred during a non-teaching period, so all focus groups were conducted online. Although studies comparing online and face-to-face focus groups report mixed findings (Jones et al., 2022), meaningful group interaction can emerge in online formats (Dedios-Sanguineti et al., 2025), and the online format provided the only practicable means of convening the participating students during the study window.

## Theoretical Framework

We framed our study using the logic of ECT (Oliver, 1980), which posits that satisfaction is determined from prior expectation, and subsequent confirmation or disconfirmation of those expectations. In other words, prior to purchasing or using a product, individuals hold expectations about its use. After actual use, the difference between the expectations and the actual performance determines satisfaction (Pan et al., 2024). ECT has been used in studying intent to use online learning platforms (Pan et al., 2024; Riyat & Kakkar, 2025) and acceptance and continuation of use in e-textbooks (Stone & Baker-Eveleth, 2013). ECT has also been used to explain university students' continuance intention towards online learning during the COVID-19 pandemic (Wang et al., 2021). We did not use ECT as a predictive model for quantitative analysis, as is often the case. Rather, we drew on the broader notion that evaluation is situated in the spaces between experience and expectation.

We developed an interview guide structured around four categories: functional, affective, normative, and identity. These categories were developed iteratively by the authors to structure elicitation and achieve a broad range of topics of discussion. To ensure coherence, the post-use questions were written as counterparts to the pre-use questions. The structure of the guide is shown in Table 1.

*Table 1: Pre- and Post-Use Interview Guide Structure*

| Category | Focus | Pre-Use Question | Post-Use Question |
|---|---|---|---|
| Functional | What the platform is expected to do and is experienced as doing. | Based on what we've told you so far, what do you think this platform is supposed to do in terms of function? | Thinking about your expectations of the platform, and what you expected it to do, to what extent did the reality match that? |
| Affective | How using the platform is expected to feel and feels. | Thinking about writing in this platform, how do you imagine that will feel, compared to your normal writing environment? | How did it actually feel to write within the platform in comparison with how you expected it to feel beforehand? |
| Normative | How the platform is expected to speak to fairness, transparency, and appropriateness. | What would this platform need to do for you to feel it was being used fairly? | Did the platform feel it was transparent about what it was doing and why? Did that match what you said you'd need for it to feel fair? |
| Identity | How the platform is expected to impact, and is experienced as impacting, authorial identity. | Do you expect the platform to affect how you think about AI use during the task, or how you decide whether to use it? | How did you decide whether to use the AI features or not? What was the decision-making like? |

**Recruitment**

Participants were enrolled students at a UK university and recruited via an open call for volunteers distributed through internal student channels (website and newsletter). Participation was voluntary and had no impact on, nor relation to, students' programme of study. No structured demographic data was collected, and contextual details were only recorded if volunteered by participants in the focus groups. Participants received a £100 voucher for taking part. The Participant Information Sheet (PIS) documented that funding for the study was provided by Turnitin, but specified that Turnitin was not involved in the design, data collection, analysis, interpretation, or publication of the findings. Induction training to the platform for the principal investigator and student participants was declined to preserve an unmediated first encounter with the platform.

Seven students participated, spanning undergraduate, postgraduate taught, and doctoral study across arts and humanities, STEM, and the social sciences. The pre-use phase comprised two focus groups scheduled separately to accommodate participants' availability (FG1A, N=4; FG1B, N=3), followed by one post-use focus group (FG2, N=7). Six participants completed the writing task. One participant (P05) encountered technical difficulties and could not access the platform; their post-use contribution is therefore treated as an account of attempted access rather than task completion, and this is acknowledged as a study limitation.

**Procedure**

Focus groups were conducted virtually via Microsoft Teams. Sessions lasted approximately 60 minutes and were video and audio recorded with consent for transcription purposes. Automatic transcription was used, followed by human verification and editing. Recordings were deleted on completion of transcription and anonymisation.

The pre-use focus groups examined expectations of the Turnitin Clarity platform. To ensure integrity of pre-use expectations, the platform was not described in detail, but a link to the platform's website, along with a brief description in the PIS, was shared with participants.

Following the pre-use focus groups, participants were enrolled onto the Turnitin Clarity platform and asked to complete a task within five days. The task consisted of producing a short, 500-word piece of writing in response to a prompt (Appendix 1). To ensure that the task completion reflected authentic participant use, we informed participants that they could use as many or as few of the AI features as they wished. The task prompt was developed to be broad, not specific to a single discipline, and on a mainstream, familiar topic:

> "Examine the impact of social media on everyday communication. Use academic sources to support your arguments and include specific examples."

The task was deliberately unassessed and had no grading rubric; participants were informed through the PIS and verbally that neither writing quality nor content would be evaluated. This design was intended to reduce perceived pressure to perform and to remove any direct or perceived academic advantage from producing a high-quality answer. The study concerned participants' experience of the platform and the process of writing within it, not the quality of the resulting product. Although the absence of grades may limit ecological validity relative to credit-bearing assessment, it allowed participants to explore the platform without believing that their performance would affect their academic outcomes. The task was open for approximately five days, and participants were able to complete the task at their own time and pace, using their own devices. The novel features of the platform, including the AI chat, were enabled for participant use. The AI chat within the platform was set to allow all possible functions, including the default (asking questions to enable students to reflect on their ideas) along with optional categories (planning and ideas, revisions, proofreading, and illustrative examples). Response complexity for the AI tool was set to 'Standard', meaning that moderate length responses would be provided with standard sentence structures and academic terms. Participants were able to view the writing reports on their work. Responses were not saved within the platform's repository, nor used for future model training, and their submissions were deleted by the provider on completion of the activity.

The post-use focus group was convened at the end of the task window and brought together members of FG1A and FG1B in a single session (N=7). This format may have influenced group dynamics. All had encountered the same study protocol, although one participant had been unable to access the platform. After the structured questions, the group was invited to reflect more broadly on the platform and its implications for educational assessment.

**Data Analysis and Reflexivity**

Transcripts were anonymised prior to analysis, and any names or identifying information were removed. Reflexive thematic analysis was undertaken, from an interpretivist and social constructionist perspective. Thematic analysis is often described as a group of different approaches which can range from "scientifically descriptive" to "artfully interpretive" (Braun et al., 2022, p. 19). The reflexive approach used here involves the active construction of themes from the dataset by the researchers, representing stories of shared patterns of meaning, rather than topic summaries. Equally, our analysis recognises that inductive and deductive and semantic and latent coding operate on a continuum (Braun & Clarke, 2019). Our deductive orientation is based in the ECT framing, in which pre- and post-use accounts were compared to identify variance or confirmation of prior expectations, while coding within the transcripts themselves remained inductive and grounded in the participants' responses. In line with this method, we did not develop codebooks or conduct inter-rater reliability measures, but instead relied on prolonged immersion, reflexivity, and engagement with the data (Braun & Clarke, 2019). Coding took place through the ATLAS.ti qualitative data analysis platform and involved an iterative and recursive process. This was based on the features described by Braun and Clarke (2021) as part of the reflexive method: reading, reflecting, questioning, and returning to the data frequently. Transcripts were analysed holistically and as a single dataset, going between pre- and post-use accounts, and themes are presented as overarching categories that were developed from the data, rather than discrete themes from each phase.

Both authors research academic integrity and generative AI and have published works that are critical of AI text detection software and its consequences. As researchers, we came to this study expecting that surveillance would be a prominent concern and that students would resist the notion of having process data recorded. This led us to initially be sensitive to accounts of discomfort and refusal and potentially made us less sensitive to accounts that made the platform appear fair, useful, or neutral. In our coding process, we deliberately aimed to find disconfirming accounts that stood against our analysis. Examples include P02's account of feeling more motivated because recording made the task seem fairer, and P07's description of becoming "more pro the platform" after using it. We report these alongside the more critical accounts.

# Results

The analysis resulted in the generation of 125 initial codes, refined to 117 after controlling for duplicates and merging similar codes. These were developed into six themes. Quotations are reproduced verbatim, with square brackets marking minor editorial clarifications, and participants are labelled as P01 to P07. FG1A and FG1B refer to the two pre-use focus groups, whereas FG2 refers to the post-use focus group. Table 2 provides an overview of the six themes before each is examined in detail.

*Table 2: Thematic Structure of Results*

| Theme | Interpretive focus |
|---|---|
| 1. The Desire for Two-Way Transparency | Being observed made writing feel like a performance for an imagined assessor, and participants wanted visibility into both what data was captured and how assessors and institutions would interpret and use process data. |
| 2. Contested Fairness as Justification | Process visibility could mitigate unequal access to AI, but fairness did not fully justify surveillance and could be undermined by circumvention. |
| 3. Writing Defensively | Fear of false accusation led participants to retain evidence, alter lexical and stylistic choices, and sometimes lower their academic register, a vigilance the platform did not remove. |
| 4. Assistance and Refused Co-Authorship | A bounded assistant supported grammar and clarified acceptable use of GenAI features, but could weaken ownership, learning, and diversity of output. |
| 5. Linear Process Capture and the Realities of Writing | A linear online record could not fully represent handwritten, iterative, multi-document, and offline writing practices, limiting perceived validity. |
| 6. Conditional Acceptance and Preconditions for Trust | Acceptance depended on practice time, two-way transparency about institutional use, and opt-in data governance, with confidence developing through experience. |

**The Desire for Two-Way Transparency**

Across both stages of the study, participants viewed the platform primarily as an instrument of observation and surveillance, which was intended to consistently monitor students without providing insight into how data might be generated and used by the observer. In the pre-use focus groups especially, anticipated experiences focused on ideas of invasiveness and surveillance:

> *"I would say this, I think invasive is like the best word for me. Like I feel like... They're seeing through my head while I'm doing stuff and I don't really like that. But I also get why it's important, but it's just a bit uncomfortable thinking about it."* (P01, FG1A)

After using the platform, this feeling still seemed to remain. One participant described the reminders of staff visibility as provoking a menacing feeling:

> *"There's one thing that I found quite uncomfortable. It was when you open the AI chat part, it says at the top something like, your lecturer will be able to see all that you've typed here, or like the amount of AI usage... having that constantly there was a bit menacing almost."* (P04, FG2)

Participants anticipated that this feeling of being observed would reshape how they engaged in writing, leading to a feeling of 'performing' for an imagined assessor rather than focusing on the writing itself:

> *"I think that the process of writing it would feel more like a performance, if that makes sense. I wouldn't be working necessarily to the best of my abilities in like in my natural way necessarily, because I would be focused on making sure it looks like I'm working in a normal way."* (P07, FG1B)

One participant also described this after using the platform:

> *"There were a lot of things that I was constantly aware of, and perhaps then I was focusing less on the task itself and more on how I was presenting my thought process throughout the task. But then that was slightly outweighed by the concept of I'm more motivated because this feels like a more fair process."* (P02, FG2)

These accounts support Fawns et al.'s (2026) argument that process-based assessment can increase performativity on the part of the student-writer. In suggesting possible solutions, participants frequently noted a need for increased transparency on behalf of the assessor and the process data they would be evaluating. This included a desire for communication on what is recorded, how data is displayed, and how it would be used:

> *"I feel like I want to see what has been actually recorded since like the whole platform is connected to the writing process and, you know, transparency is kind of important here. So I feel like if I could sort of see how they record my whole process. I think I will understand it better and perhaps I will feel more assured as well."* (P06, FG2)

Participants found that the platform transparently declared what data would be collected, as there was a pop-up list the first time they logged into the platform, but they found it difficult to understand what institutional policies would be enacted based on the data, and what assessors would do with this data. This concept of institutional transparency is also featured in Theme 6.

**Contested Fairness as Justification**

When discussing fairness and transparency, participants described process visibility as having the potential to reduce inequities created by uneven access to AI tools. In response to the pre-use question, 'What do you think this platform is supposed to do in terms of function?', one participant explained:

> *"To enable AI use to occur in a way that's fair and equal for everyone, because currently it's dependent on who has access to what platform, who's paid what."* (P02, FG1A)

In the post-use focus group, it seemed that this promise of fairness was a necessary justification for the intrusiveness and obstructiveness of the platform for this participant:

> *"Maybe that's something I'd get used to in future, but I know with this one as the first step, it did get in the way. But then it felt like a necessary get in the way because I was more, I know if this had been one that counted, I would have been more motivated to do it because it felt fairer because at least I knew everyone was being recorded."* (P02, FG2)

On the other hand, not all participants shared this view that the platform's data collection was a price worth paying for fairness. For one participant, the prospect of the platform's process data collection was a point of refusal:

> *"I wouldn't go, honestly, if I, if I was, if I was a young person or even like an adult, like I'm a mature student and I, and I, I just wouldn't go, I wouldn't go to university if I knew that that was going to be the full process, it would put me off."* (P05, FG1B)

This demonstrates the subjective, varying perspectives on what constitutes an acceptable trade-off between fairness and surveillance. Participants also viewed fairness in terms of their position temporally, as members of a cohort that are part of a major disruption: studying at a time while higher education adapts to new, post-GenAI forms of learning and assessment:

> *"I don't feel like anyone should be at a disadvantage for being the guinea pigs kind of thing."* (P05, FG1B)

Despite views that the platform could enhance fairness, there was also recognition that strategies to evade the platform using AI could still be used to bypass process capture, as shown in this interaction between P05 and P06:

> *"Yeah, so if you just wrote it somewhere with AI and then copied and pasted that into a document, it can't track anything, can it? Is that what, that's what it's, is that how you read it?"* (P05, FG1B)

> *"No, I was just thinking because from what I understood is that Turnitin Clarity would see how you produce the whole piece of writing from the beginning of the draft or how you actually edit every single note, every single word. And I could have made that up by having a complete written by AI, by maybe by ChatGPT, and I just take my time to write it into the document."* (P06, FG1B)

**Writing Defensively**

Fernandes and McIntyre (2025) highlighted cases of students collecting evidence to proactively defend themselves from accusations of misconduct. A related practice, managing the surface features of a text so that it reads as recognisably human, has been described elsewhere as part of a performance of legitimacy (Roe, Perkins, Bannister, et al., 2026). Our focus groups suggested that participants are acutely aware of the risks of a false accusation of misusing AI, and several participants already moderated their writing with this in mind. This feeling is captured clearly by P05:

> *"You know that feeling you get when you're driving down the road and a police car comes behind you and you're doing absolutely nothing wrong, but you feel like you are because there's a police car behind you. I kind of feel like that's what Turnitin makes you feel like."* (P05, FG1B)

Other participants explained how they had already sought to guard against AI accusations by retaining evidence or altering their writing:

> *"Currently I just have stacks of notes that I've kept just in case somebody is like, it's AI generated. I'm like, no, I promise it's not. I've got all these notes. And so whether I'd still have to do the same for the platform would be maybe a concern that I'd have."* (P02, FG1A)

> *"It's like, oh, I can't use this em dash because if I use it, then it's like a telltale sign of AI. Or if I use like the words like quote unquote 'delved', then that's also a telltale sign of AI, like so I can't use that word."* (P03, FG1A)

In relation to this, some participants speculated that Clarity could help automatically store evidence that could help defend them from accusations of AI misuse:

> *"Like if I submitted a piece of work that was written on a platform where the marker had pretty much like access to every single line I typed, I would feel like, okay, I've done nothing wrong then. Like, so even if I'm accused, then there's like already evidence that, you know, like I didn't use AI, I didn't generate this entire essay. Like it came from my thoughts and it's my original work, essentially."* (P03, FG1A)

For P07, defensive writing involved lowering their academic register rather than distancing themselves from the work: the essay still felt recognisably their own, but process surveillance heightened their concern that source-based complexity might be misread as copied text:

> *"For me, it definitely felt like my work. I think that it felt not as good as what I could usually produce. It felt more like my own voice rather than like an academic piece that I had written. And I think that's because there was an element of paranoia that if I had research[ed] something like quite intensely and I'm writing all of these really complex thoughts that other people have had it might flag it for plagiarism. It might think that I've copied and pasted that. I might, I might write it out too similarly to the sources and not realise it. And I consciously know that that can happen, you know, whether using this platform or not. But something about the fact that it's watching my thought process, I think made me a bit more paranoid. So I think that I compensated by writing things to a lower academic standard and it felt more like just my internal monologue putting on paper."* (P07, FG2)

In the post-use focus group, P05, who had been unable to access the platform, anticipated a similar self-monitoring that would reach into composition itself, and linked it to how far they trusted whoever would be reading the work:

> *"It feels like I have to double check what I'm writing down before I actually write it. Maybe in a good way, but perhaps it's just more daunting to... even put any random thoughts down there depending on how much you trust the person is going to rate it."* (P05, FG2)

**Assistance and Refused Co-Authorship**

The integrated AI assistant within the platform was evaluated positively in some respects, although views were mixed. It was described as useful for helping with grammar, catching errors that Word's own checker had missed:

> *"It was helpful as well with grammar checks because I had, I very much continued to copy and paste everything across into Word as like a double backup. And it caught things that Word hadn't in grammar. So that was helpful."* (P02, FG2)

Participants also viewed it as appropriate that the AI refused to assist with generating long passages of text, or independently research topics. This drawing of a line of 'acceptable use' of AI was seen positively:

> *"I think its self-limitations did help a lot and made me more confident in using it because I knew it wouldn't do something the instruction said you can't use it for."* (P04, FG2)

Where the assistant did contribute more than just grammar and structural checks, some participants felt that it impacted their feeling of authorial ownership and negatively affected learning:

> *"But having it as a tool and using it the way that I did meant that the AI did all of this for me. So I ended up actually not really remembering what I wrote, if that makes sense. So I felt like I didn't really learn that much about the topic of the essay, but the task itself, I guess it was helpful."* (P03, FG2)

> *"I'm less proud of the aspects that it had pointed out to me... it did just feel like shortcut, like corner cutting."* (P02, FG2)

Participants reflected on the broader impact of AI assistance, considering that shared prompts and a bounded AI assistant might lead to homogeneous work, and whether existing rules of plagiarism could be applied:

> *"I can't help but wonder if everyone that's supposed to do this assignment uses this chat bot and uses this exact same given prompt to come up with a structure, then would everyone end up writing the same thing maybe? Then would this count as plagiarism?"* (P03, FG1A)

Another participant similarly described the results of the AI assistant as overly generic:

> *"I think the AI got in the way a bit with it. When I'd be using it, I'd say ask it to give me some ideas for titles when it was just very generic and then asking for like ideas on what I can write about. Again, it's just quite generic. And I feel like if I was like dedicated to using this AI to help me, I just get a very bland essay in the end."* (P04, FG2)

**Linear Process Capture and the Realities of Writing**

While the platform records the writing process, participants expressed concern that it could not capture the various ways in which they write. Several participants documented that they rely on multiple documents, handwritten notes, iterative 'messy' notetaking, and thinking about work while away from the computer, and held concerns that this would not or could not be documented:

> *"I hand write everything kind of first because I think better somehow. And so I would maybe be worried that that wouldn't translate onto the platform or would I have to take photos of this to prove the planning process?"* (P02, FG1A)

A deeper worry related to the ability for the platform to interpret this messy nature of planning and writing:

> *"I feel like I'm also more aware that my writing is quite messy and like I pause a lot of times, I delay, I rewrite, a lot of these things that people do. And I don't know if the system would actually interpret that correctly."* (P06, FG2)

Reliability of access compounded this concern, with some noting that when they lost their connection to the internet or were writing while on public transport and the connection dropped, their work would stall. Before using the platform, one participant had suggested that it might suit specific assessment types, especially bounded, timed assessments, reasoning that AI use in such assessments was otherwise impossible to control:

> *"I think that the 24-hour online exams that the university do, I think it's a perfect place for it because a lot of students do use AI anyway, even though it is like strictly forbidden and there's no way to control for that."* (P07, FG1B)

**Conditional Acceptance and Preconditions for Trust**

After the usage period, no participant rejected the platform outright, but none accepted it unconditionally. Acceptance of the platform was described as conditional and based on several different facets. The first of these was habituation, familiarity, and unstructured practice time to develop an understanding of the platform:

> *"The platform kind of describes what the platform does, which is great for understanding, but I'd say the students also need to be aware of what the lecturers, what the staff are going to do with that, because and how that's going to be factored into account in the assessment. And that would be definitely down to the individual department. So there's still a lot of need for a lot of clarity. And then also to factor in like a training time slot, kind of similarly for the platform itself, completely unassessed where you can just play around with it and try the different tools. Because if an assignment is the first time you've experienced this platform, there's a lot of processing learning time, even though it is quite similar to existing platforms."* (P02, FG2)

A second important concern was data governance. Participants wanted the option to opt in specifically for the use of their process data, and minimal use and retention of process data. In response to a question about who should have access to process data, one participant stated:

> *"I think I would hope that it's only the lecturer or whoever's marking it would have access to it. And if Turnitin wanted to use it for any other things, it would be an opt-in thing. It would ask you, are you okay with this being given to someone else or used for any sort of AI training? I wouldn't want that done by default, and you have to use that."* (P04, FG1A)

Across the study, most participants felt that the platform matched their initial expectations:

> *"I don't think it really changed that much for me. I still think it's just a way to monitor student work and like give everyone access to AI. Yeah, that's, it hasn't really changed even after I did it."* (P01, FG2)

Some participants also felt more accepting of it after being able to use it for writing themselves:

> *"For me, having used the platform, I think I'm much less, I guess the right word is less sceptical of it, because I feel like I understand why it's there, what it does, what it's for. And I'm a bit less like, why would we be using this? I get it, basically, I understand its purpose and I think it has a good purpose and I think it does its job. So I'd say I'm more pro the platform now."* (P07, FG2)

## Discussion

This study aimed to explore how university students' pre-use expectations of Turnitin Clarity compared with their actual experiences, and what this suggested about acceptability in higher education assessment. Drawing on ECT (Oliver, 1980), the overall finding is one of mixed confirmation. Participants largely felt that the functions of the platform were as they expected, consisting of a monitoring platform with a bounded AI assistant. Affective expectations were also largely confirmed: the sense of being 'watched' that participants had anticipated persisted after use, as Theme 1 shows. For some, it was tempered by the perception that recording everyone made the process fairer; for others, it heightened self-monitoring.

On the practical side, participants also felt that some features, such as the AI assistant and the requirement of internet connectivity, fell below their expectations. Overall, participants weighed perceived gains in fairness and equality against potential costs to their workflow, learning quality, and epistemic agency.

Several participants did not outright resist the idea of having their writing process documented, while others expressed anxiety, in some cases "extreme anxiety" (P05, FG1B). This was at times counterbalanced by the perception that surveillance could enhance fairness and mitigate the unequal advantages resulting from AI use. Participants' perspective on acceptability also required reciprocity. Participants were highly concerned with the lack of two-way information they felt the platform could present.

Participants brought pre-existing defensive writing habits to the platform, gained from a fear of accusation of incorrectly using GenAI tools. Before they had used Clarity, they described avoiding punctuation and lexical choices associated with machine-generated prose, retaining handwritten notes as precautionary evidence, and anticipating the possibility of an AI-misuse accusation. Writing within the platform did not remove this vigilance and, for some participants, added a further layer to it. Although Clarity recorded process evidence that could support an authorship claim, P07's account suggests that this did not create a secure space for ordinary academic writing: surveillance led to further defensive adaptation, and influenced the participant to simplify their register because of the perception that complex, source-based writing might be misread as copied text. P05, who had been unable to access the platform, anticipated similar effects, describing a need to "double check" what they wrote before writing it and finding it "more daunting to... even put any random thoughts down". This interpretation aligns with evidence from Adnin et al. (2025) that students may restrict disclosure of GenAI use, even where it may be acceptable, because they anticipate negative judgement about authenticity, effort, or dependence. For P07, then, the platform did not only record a pre-existing writing process; it took part in reshaping it further. What Clarity captured, on this account, was writing already adapted to the expectation of being read for signs of AI use, then adapted again to the fact of being watched.

In relation to the AI assistant, participants accepted that it should be bounded and limited. For some, those boundaries clarified the line between acceptable and unacceptable use and reduced uncertainty, consistent with analyses of institutional guidance and student interpretations of acceptable AI use (An et al., 2025; Corbin et al., 2025; Perkins & Roe, 2024). This suggests that the assistant can perform normative as well as structural labour. However, even a bounded assistant could weaken feelings of authenticity, authorship, and learning, echoing concerns about cognitive offloading (Bastani et al., 2025; Gonsalves, 2026). Strömberg et al.'s (2026) Chinese secondary-education study further found that access to generative AI increased homework performance while reducing subsequent examination performance, indicating a possible learning penalty when assistance substitutes for cognitive effort. Participants also worried that shared prompts and a common assistant would produce homogeneous, less diverse work, in line with evidence that generative AI can increase individual creativity while reducing collective diversity (Doshi & Hauser, 2024), and with arguments that GenAI risks algorithmic homogenisation (Roe, Furze, & Perkins, 2026).

Many participants shared the feeling that the platform only captured one facet of data in an otherwise complex, multimodal writing process. The word 'messy' was repeatedly used to describe the process of composing written assessed work. Using a single process document as a proxy for assessing the entire writing process was of concern to students, many of whom worked on multiple documents, both digital and physical, and maintained that a lot of their thinking, planning, and preparing could not easily be documented in the Clarity platform. This led to a concern that assessors would not be able to gain a valid judgement on their writing process. This suggests that process-capturing platforms need space for further documentation, including the option to include photos of handwritten notes, reflections, or other multimodal data. Bounded, timed assessments may be less affected by this issue; one participant suggested that 24-hour online exams would be a "perfect place" for the platform, although their reasoning concerned the control of AI use.

### Implications

This study has several practical implications for the use of process-documenting writing platforms. First, participants in this study expressed a strong need for two-way transparency, including clear information on what data would be captured, the standard it would be held to, and being able to view the marker-facing user interface for themselves. Second, participants indicated a critical need for proper training, unassessed time to practise with the platform, and space to familiarise and habituate to it. Third, participants noted that having explicit, clear data governance, opt-in rather than opt-out procedures for secondary use of data, and clear deletion timelines is required. These preconditions recurred across both phases of the study.

Process evidence may also affect marking and feedback workload. Winstone et al. (2026) argue that any increase in this burden can be minimised if process evidence is tightly aligned with learning outcomes and presented through concise, structured 'windows', such as annotated drafts or brief reflections. Institutions should therefore avoid adding an unbounded second product for markers to assess; instead, they should rebalance final-product requirements, specify which process traces warrant attention, and use rubrics that support selective review.

### Limitations

This is a small-scale study (N=7) across a single institution. Participants were self-selected, and the task was unassessed. While the study construct remains suitable for addressing the research questions, the findings may diverge when the platform is used for assessed work, under graded conditions. One participant was unable to access the platform, which limited their ability to participate in the post-use focus group fully. A further limitation is that the study focused specifically on student perspectives, leaving the question of how markers experience the platform, as well as use and interpret process data, open for future research. The study was funded by Turnitin, the provider of the platform being studied, and Turnitin provided access to the platform free of charge. Turnitin had no role in the design, analysis, or reporting, but readers should bear this relationship in mind when interpreting the findings. Finally, onboarding and induction training to the platform was deliberately declined, to preserve an unmediated encounter with Turnitin Clarity. This was a considered trade-off, as it allowed pre-use expectations to be viewed in relation to unguided experience. At the same time, it means that some of the difficulty or challenges that participants faced could result from the lack of induction, rather than the platform's user experience itself.

## Conclusion

This exploratory study provides one of the first empirical accounts of students' pre-use expectations and post-use experiences of a process-capturing writing platform with a bounded AI assistant. Although the sample was small, the longitudinal focus-group design revealed a more complex pattern than straightforward acceptance or rejection of being observed. Across the six themes discussed, participants weighed the possible benefits of fairness, evidentiary protection, and bounded assistance against surveillance, workflow disruption, learning loss, and uncertainty about how process data would be judged.

Direct experience made some participants less sceptical and helped them understand why process evidence might be useful. Yet the protection promised by the platform did not remove defensive writing. Before using the platform, participants already described altering punctuation, vocabulary, and evidentiary practices to avoid being perceived as using GenAI. After using it, some described further self-monitoring, including lowering their register so that source-based writing would not be read as copied, as the visibility of their process introduced a further pressure to perform an acceptable form of authorship. Process capture therefore does not act as a neutral observer of student writing. It reorganises the very process it sets out to record, and any inference drawn from that record must be read in that light. Further research should test these conditions in assessed settings, with larger and more diverse samples, and from the perspectives of markers as well as students.

# Disclaimers

### Declaration of Generative AI and AI-assisted technologies in the writing process

GenAI tools were used for ideation and in some passages of draft text creation, which was then substantially revised, along with editing, formatting, and refinement during the production of the manuscript. The tools used were ChatGPT (Sol 5.6) and Claude (Sonnet 5, Opus 5, 5.5), which were chosen for their ability to provide sophisticated feedback on textual outputs. These tools were selected and used supportively and not to replace core author responsibilities. The authors reviewed, edited, and take full responsibility for all outputs of the tools used.

### Funding Statement

This study received funding from Turnitin LLC, which also provided free access to the Turnitin Clarity platform. Turnitin had no input into the design of the study, collection, analysis or interpretation of data, writing of the manuscript, or decision to submit the manuscript for publication.

### Conflict of Interest Statement

Beyond the Turnitin funding and platform access declared above, the authors declare no conflicts of interest.

# Appendix 1: Writing Task Instructions

*This writing task formed part of the study and was supplied to participants in full.*

This writing task forms part of a research project exploring student experiences of AI-enabled writing and assessment platforms. The task is intended to provide you with hands-on experience using the Turnitin Clarity platform and is not assessed. The quality of your writing will not be graded and participation will have no impact on your university studies, modules, or academic records.

You are asked to write approximately 500 words in response to the following prompt: “Examine the impact of social media on everyday communication. Use academic sources to support your arguments and include specific examples.” Your response should present a clear discussion or argument, engage with relevant evidence or literature, and be written in an academic style.

You can adopt any perspective on the topic provided your discussion is supported with explanation and evidence. You should complete the task individually within the Turnitin Clarity platform.

You may engage with any of the features available within the platform. Use of AI tools is optional and there is no expectation that you use AI assistance during the task. The purpose of the activity is to explore your experiences of writing within the platform rather than assess writing quality or patterns of AI use.

The task may be completed at any point between the first and second focus group sessions. There is no grading rubric or formal assessment criteria attached to the activity. Its sole purpose is to support discussion and reflection.